\documentclass[twocolumn,aps,pre,epsf,superscriptaddress]{revtex4-2}

\usepackage{graphicx}
\usepackage{dcolumn}
\usepackage{bm}
\usepackage{siunitx}
\usepackage{soul}
\usepackage{color,ulem}
\usepackage{float}
\usepackage{stfloats}
\usepackage[section]{placeins}
\usepackage{amsmath}
\usepackage{amssymb}
\usepackage{epsfig}
\usepackage[T1]{fontenc}
\usepackage[utf8]{inputenc}

\begin{document}

\title{Critical dynamical behavior of the Ising model}

\author{Zihua Liu}
\email{z.liu1@uu.nl}
\affiliation{Department of Information and Computing Sciences, Utrecht University, Princetonplein 5, 3584 CC Utrecht, the Netherlands}

\author{Erol Vatansever}
\email{erol.vatansever@deu.edu.tr}
\affiliation{Department of Physics, Dokuz Eyl\"{u}l University, TR-35160, Izmir, Turkey}

\author{Gerard T. Barkema}
\email{G.T.Barkema@uu.nl}
\affiliation{Department of Information and Computing Sciences, Utrecht University, Princetonplein 5, 3584 CC Utrecht, the Netherlands}

\author{Nikolaos G. Fytas}
\email{nikolaos.fytas@coventry.ac.uk}

\affiliation{Department of Mathematical Sciences, University of Essex, Colchester CO4 3SQ, United Kingdom}

\date{\today}

\begin{abstract}

We investigate the dynamical critical behavior of the two- and three-dimensional Ising model with Glauber dynamics in equilibrium. In contrast to the usual standing, we focus on the mean-squared deviation of the magnetization $M$, MSD$_M$, as a function of time, as well as on the autocorrelation function of $M$. These two functions are distinct but closely related. We find that MSD$_M$ features a first crossover at time $\tau_1 \sim L^{z_{1}}$, from ordinary diffusion with MSD$_M$ $\sim t$, to anomalous diffusion with MSD$_M$ $\sim t^\alpha$. Purely on numerical grounds, we obtain the values $z_1=0.45(5)$ and $\alpha=0.752(5)$ for the two-dimensional Ising ferromagnet. Related to this, the magnetization autocorrelation function crosses over from an exponential decay to a stretched-exponential decay. At later times, we find a second crossover at time $\tau_2 \sim L^{z_{2}}$. Here, MSD$_M$ saturates to its late-time value $\sim L^{2+\gamma/\nu}$, while the  autocorrelation function crosses over from stretched-exponential decay to simple exponential one. We also confirm numerically the value $z_{2}=2.1665(12)$, earlier reported as the single dynamic exponent. Continuity of MSD$_M$ requires that $\alpha(z_{2}-z_{1})=\gamma/\nu-z_1$. We speculate that $z_{1} = 1/2$ and $\alpha = 3/4$, values that indeed lead to the expected $z_{2} = 13/6$ result. A complementary analysis for the three-dimensional Ising model provides the estimates $z_{1} = 1.35(2)$, $\alpha=0.90(2)$, and $z_{2} = 2.032(3)$. While $z_{2}$ has attracted significant attention in the literature, we argue that for all practical purposes $z_{1}$ is more important, as it determines the number of statistically independent measurements during a long simulation.

\end{abstract}

\maketitle

\section{Introduction\label{sec_intro}}

Universality is a key concept in statistical physics~\cite{fisher74}. Phenomena which at a first glance seem completely unrelated, such as the liquid-gas phase transition and the ferromagnetic-paramagnetic phase transition in magnetic materials, belong to the same universality class, sharing the same set of critical exponents and other renormalization-group invariants that characterize their equilibrium behavior around the critical point~\cite{fisher98}. The Ising model~\cite{Ising25}, the simplest fruit-fly model in statistical physics which lends itself well for theory and simulation, is found to belong to the same universality class~\cite{landau_book,barkema_book,amit_book}. Studies of the critical equilibrium properties of the Ising model are therefore of direct experimental relevance~\cite{landau_book}.

The concepts of critical phenomena can fortunately be extended to dynamical processes -- for a seminal review see Ref.~\cite{hohenberg77}. However, while universality is well established for equilibrium properties, it is not clear in how far it also extends to dynamical properties~\cite{hohenberg77,folk06,hasenbusch07,zhong20}. As it is well-known, the onset of criticality is marked by a divergence of both the correlation length $\xi$ and the correlation time $\tau$. While the former divergence yields singularities in static quantities, the latter manifests itself notably as critical slowing down. To describe dynamical scaling properties, an additional exponent is required in addition to the static exponents. This so-called dynamic exponent $z$ links the divergences of length and time scales, i.e., $\tau \sim \xi^{z}$~\cite{nightingale96,hasenbusch20}. In a finite system, $\xi$ is bounded by the linear system size $L$, so that $\tau \sim L^{z}$ at the incipient critical point.
The dynamic critical exponent $z$ has been numerically computed to be $z = 2.1665(12)$ at two dimensions by Nightingale and Bl\"ote ~\cite{nightingale96}. Note the value $z = 2.0245(15)$ at three dimensions~\cite{hasenbusch20}. 
 
 In the current paper we attempt to extend our knowledge in the field by highlighting an overlooked aspect of dynamic critical phenomena using single spin-flip (Glauber) dynamics on the two- and three-dimensional Ising ferromagnet. In contrast to the standard belief that the dynamical critical behavior is characterized by a single dynamic exponent $z$, we provide numerical evidence that there is another dynamic critical exponent, considerably smaller than the most studied one, which appears to be of greater practical relevance. In particular, we provide a more refined description of the magnetization autocorrelation function featuring three regimes that are separated by two crossover times, namely $\tau_{1}\sim L^{z_{1}}$ and $\tau_{2}\sim L^{z_{2}}$, where $z_{1}$ is a newly identified dynamic exponent and $z_{2}$ the already well-known exponent~\cite{hasenbusch07,zhong20,nightingale96,hasenbusch20}. 

The rest of the paper is laid out as follows: In Sec.~\ref{sec_model} we introduce the model and outline the numerical details of our implementation. In Sec.~\ref{sec_results} we introduce the key observables under study and elaborate on the analysis of the numerical data, placing our findings into context. Finally, in Sec.~\ref{sec:summary} we critically summarize the main outcomes of this work in the framework of the current literature and also set an outlook for future studies.

\section{Model and Numerical Details}
\label{sec_model}

We consider the nearest-neighbor, zero-field Ising model with Hamiltonian 
\begin{equation}\label{eq:ising_hamiltonian}
	\mathcal{H} = -J \sum_{\langle i,j \rangle} \sigma_i \sigma_j,
\end{equation}
where $J > 0$ indicates ferromagnetic interactions, $\sigma_i = \pm 1$ denotes the spin on lattice site $i$, and $\left\langle \ldots \right\rangle$ refers to summation over nearest neighbors only. Here, we study the two- and three-dimensional Ising model on the square ($L \times L$) and simple cubic ($L \times L \times L$) lattices respectively, employing periodic boundary conditions. Many equilibrium properties of these models are known, especially at two dimensions where exact results are available, such as the location of the critical temperature, i.e., $T_{\rm c} = 2 / \ln \left( 1+\sqrt{2} \right) = 2.269185\ldots$~\cite{onsager44}. For the three-dimensional model on the other hand, there is a wealth of high-accuracy estimates of critical parameters from various approximation methods, see Ref.~\cite{kos16} and references therein. One such prominent example is the value of the critical point $T_{\rm c} = 4.511523\ldots$, recently proposed in Ref.~\cite{ferrenberg18} via large-scale numerical simulations. 

The Ising model is without doubt a prototypical model for studying dynamical properties. For this purpose, an elementary move is a proposed flip of a single spin at a random location, which is then accepted or rejected according to the Metropolis algorithm~\cite{metropolis53}. One unit of time then consists of $N = L^2$ elementary moves at two dimensions (similarly, $N = L^3$ at three dimensions). This dynamics is often referred to as Glauber dynamics~\cite{martinelli99,randall00,coulon04}, even though Glauber originally used a slightly different acceptance probability. Note that transition rates in Glauber dynamics are never higher, but always at least half of those of single spin-flip Metropolis dynamics, so that all dynamic exponents are shared. Other commonly used dynamical algorithms in the extensive literature are the spin-exchange (Kawasaki) dynamics~\cite{grandi96,smedt03,godreche04}, as well as numerous types of cluster algorithms~\cite{coddington92,rieger99,bloete02}. Yet, these are outside the scope of the current work.

On the technical side, our numerical simulations of the Ising model were performed at the critical temperature~\cite{onsager44,ferrenberg18} using single spin-flip dynamics and systems with linear sizes within the range $L = \{16 - 96\}$ at two dimensions (accordingly, $L \in \{10 - 40\}$ at three dimensions). We note that the simulation time needed for a single realization on a node of a \textit{Dual Intel Xeon E5-2690 V4} processor was 1 hour for $L = 96$ at two dimensions. The analogous CPU time was 35 minutes for $L = 40$ at three dimensions. For each system size $L$, $10^4 - 10^5$ independent realizations have been generated at  both dimensions.

\begin{figure}[H]
	\includegraphics[width=8.5cm]{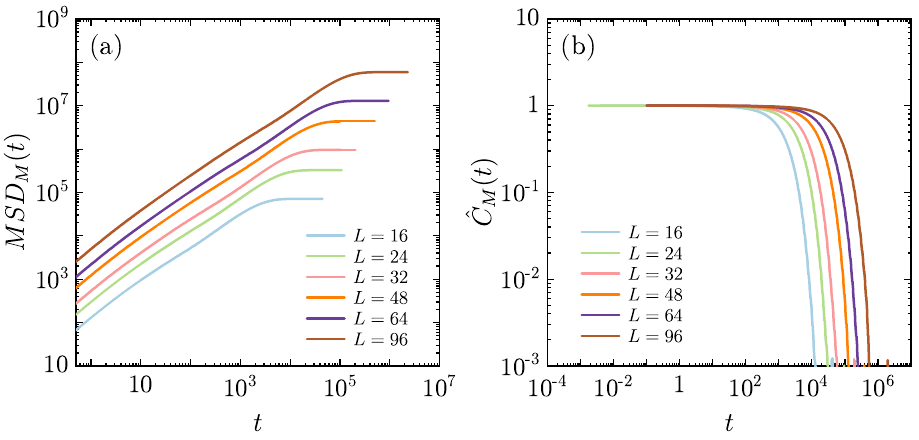}
	\caption{\label{2d1}
		(a) Mean-square displacement of the magnetization $\langle \Delta M^2(t) \rangle$ vs. time $t$. (b) The normalized autocorrelation $\hat{C}_M(t)=\langle M(t)M(0) \rangle/ \langle M^2(0) \rangle$ as a function of $t$. Results for the two-dimensional Ising model.}
\end{figure}

\begin{figure}[H]
	\includegraphics[width=8.5cm]{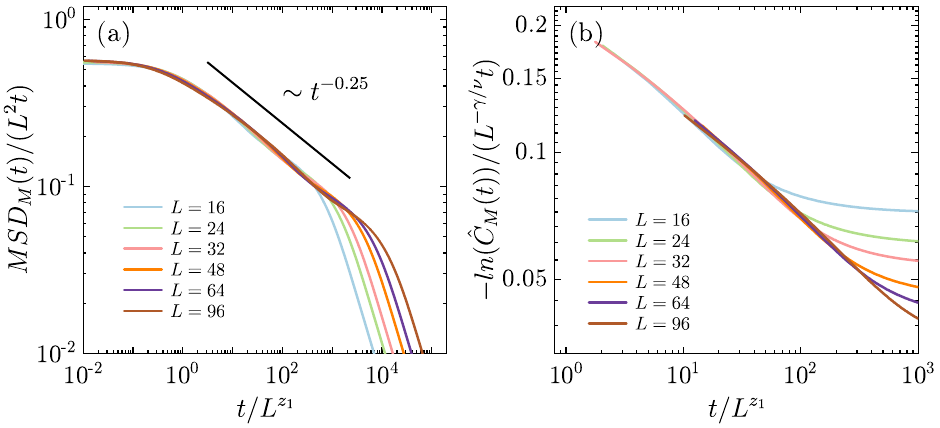}
	\caption{\label{2d2}
		(a) Data collapse of MSD$_M(t)$ curves over various system sizes around the first crossover $L^{z_1}$, with a scaling form of $\text{MSD}_M(t)/(L^2 t) \sim t/L^{z_1}$, where $z_1$ is $0.45\pm0.05$. $\text{MSD}_M(t)$ turns over from the normal diffusion ($\sim L^2t$) to anomalous diffusion ($\sim L^{2+z_1-\alpha z_1}t^\alpha$) at $t = L^{z_1}$. (b) Data collapse for $-\ln{(\hat{C}_M(t))}$ over various $L$ around $t=L^{z_1}$ with a scaling factor $L^{-\gamma/\nu}$ (note that $\gamma/\nu = 1.75$ for the two-dimensional Ising model). $\hat{C}_M(t)$ shifts from exponential to stretched exponential around $t = L^{z_1}$. Results for the two-dimensional Ising model.}
\end{figure}

\section{Results and Analysis}
\label{sec_results}

The two key observables that allow us to elaborate on some new aspects of the dynamical behavior of the Ising ferromagnet are based on the order parameter (magnetization) of the system
\begin{equation}
	\label{eq:magnetization}
	M = \sum_i \sigma_i.
\end{equation}
The first is the mean-squared deviation of the magnetization
\begin{equation}
	\text{MSD}_M(t)=\langle (\Delta M(t))^2 \rangle = \langle (M(t)-M(0))^2 \rangle,
	\label{eq:msd}
\end{equation}
and the second the magnetization's autocorrelation function, defined as
\begin{equation}
	C_M(t)=\langle M(t) \cdot M(0) \rangle.
	\label{eq:autocor}
\end{equation}

We start the presentation with the two-dimensional Ising model and the raw numerical data, as shown in Fig.~\ref{2d1}. In particular Fig.~\ref{2d1}(a) depicts the MSD$_M(t)$, whereas Fig.~\ref{2d1}(b) the normalized autocorrelation $\hat{C}_M(t)=\langle M(t)M(0) \rangle/ \langle M^2(0) \rangle$, both as a function of time. Three distinct regimes can be identified, separated by two crossover correlation times, $\tau_1$ and $\tau_2$.

At short times $t$, the dynamics consist of $L^2 t$ proposed spin flips at 
spatially separated locations, of which a fraction $f\approx 0.14$ is accepted, as determined numerically. 
The dynamics thus involve $f L^2 t$ uncorrelated changes of $\Delta M=\pm 2$. Consequently, MSD$_M$ in the short-time regime is given by
\begin{equation}
	\text{MSD}_M=4f L^2 t \quad (t \ll \tau_1).
\end{equation}
At these short times, the magnetization does not have enough time to change significantly. Hence, it stays close to its value at $t = 0$. 
The expectation of the squared magnetization is related to the magnetic susceptibility~\cite{barkema_book}
\begin{equation}
	\chi=\frac{\beta}{L^2} \langle M^2 \rangle.
\end{equation}
Thus, in the short-time regime,
\begin{equation}
	C_M(t) \approx k_b T L^2 \chi \sim L^{2+\gamma/\nu} \quad (t \ll \tau_1).
\end{equation}
Here, we used the equilibrium property $\chi \sim L^{\gamma/\nu}$.

On the other hand, at very long times the two values of the magnetization are uncorrelated so that $\langle M(t)  \cdot M(0) \rangle$ is small as compared to $\langle M^2 \rangle$. Hence we can derive that MSD$_M$ saturates as follows
\begin{equation} \label{eq:saturate}
\begin{split}
\text{MSD}_M(t) & = \langle M(t)^2 + M(0)^2 - 2 M(t)M(0) \rangle \\
 & \approx 2\langle M^2 \rangle \approx 2 k_b T L^2 \chi.
\end{split}
\end{equation}

\begin{figure}[H]
	\includegraphics[width=8.5cm]{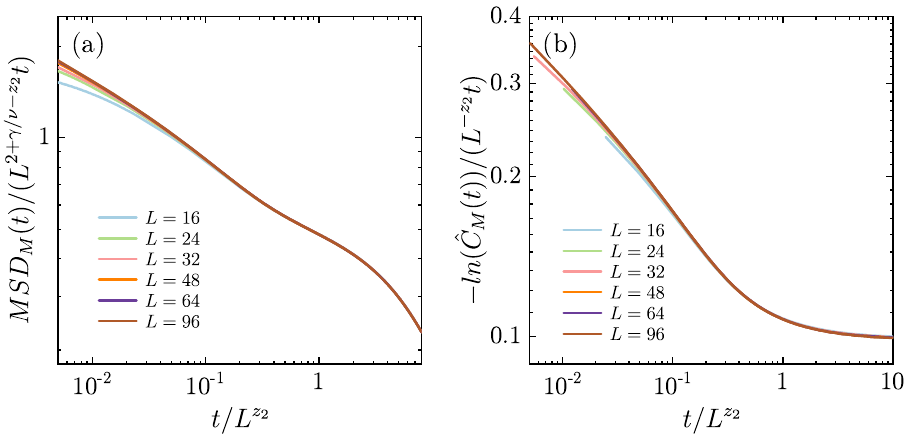}
	\caption{\label{2d3}
		(a) Data collapse of MSD$_M(t)$ curves at the second crossover $t\approx L^{z_2}$, with a scaling form of $\text{MSD}_M(t)/(L^\lambda t) \sim t/L^{z_2}$, the numerically found $\lambda$ and $z_2$ are $2+\gamma/\nu-z_2$ and $2.1667$, respectively. $\text{MSD}_M(t)(t)$ gradually transforms to saturation ($\sim L^{2+\gamma/\nu}$) from the anomalous diffusion ($\sim L^{2+z_1-\alpha z_1}t^\alpha$). (b) Data collapse for $-\ln{(\hat{C}_M(t))}$ around $t = L^{z_2}$, where the scaling factor $L^{-z_2}$ leads to an excellent collapse. $\hat{C}_M(t)$ is expected to turn over from stretched exponential to exponential around $t=L^{z_2}$. Results for the two-dimensional Ising model.}
\end{figure}

\begin{figure}[H]
	\includegraphics[width=8.5cm]{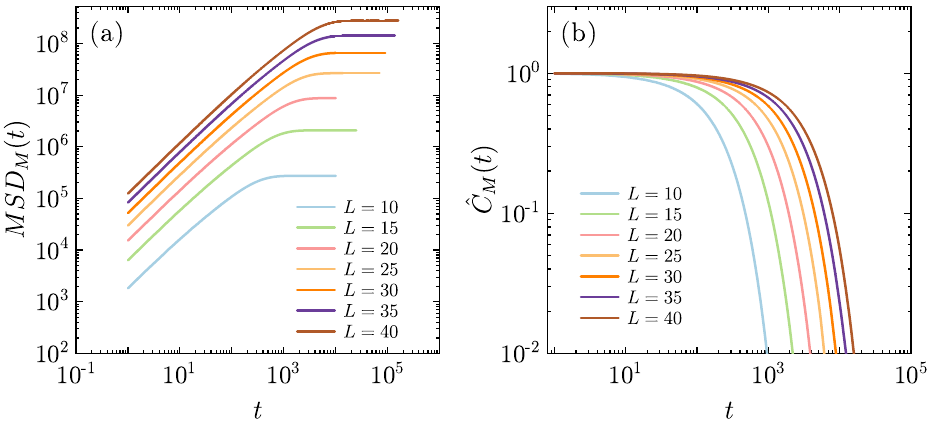}
	\caption{\label{3d1}
		Similar to Fig.~\ref{2d1} but for the three-dimensional Ising model.}
\end{figure}

Rather than an operational procedure, the dynamics can also be formulated as the application of the transition matrix $\mathcal{A}$ to a state vector $\vec{S}$. This is a rather unpractical formulation as $\mathcal{A}$ is a sparse matrix of size $2^{L^2} \times 2^{L^2}$, but nevertheless useful for the sake of argument. This transition matrix has an eigenvalue of $e_{0} = 1$, with an eigenvector in which each element lies the likelihood of that state (the Boltzmann distribution). 
It also has a second-highest eigenvalue $e_{1} \approx 1$, which determines the ultimate exponential decay of the autocorrelation. At long times $t$, the dynamical matrix is applied $t L^2$ times. Thus, expressed in $A$ the dynamics can be written as
\begin{equation}
	C_M(t) =\langle \vec{S}_t \mathcal{A}^{tL^2} \vec{S}_0 \rangle.
\end{equation}
For long times, the decay of the autocorrelation function is dominated by the largest non-zero eigenvector and eigenvalue
\begin{equation}
	C_M(t) \sim e_1^{tL^2} \sim \exp{[-t/\tau_2]},
\end{equation}
in which $(\tau_2)^{-1} = -L^2 \ln{(e_1)}$. 
It is very hard to obtain $\tau_2$ via $e_1$ numerically unless $L$ is a very small number, but this provides a valid argument to show that the magnetization autocorrelation function will decay exponentially at long times for finite $L$. Let us point out here that at times between $\tau_1$ and $\tau_2$ many modes contribute and the sum of their exponential is well-approximated by the stretched-exponential function.

As it is natural, the intermediate regime has to connect the short- and long-time regimes monotonically. The numerical data suggest that this happens via anomalous diffusion, i.e., $\text{MSD}_M \sim t^\alpha$, whereas the autocorrelation function seems to decay as a stretched-exponential with the same anomalous exponent $\alpha$.

Clearly, the key quantities that we want to establish in this manuscript are the dynamic exponents $z_1$ and $z_2$, as well as the anomalous exponent $\alpha$. To this end, we use the method of finite-size scaling~\cite{landau_book,barkema_book,amit_book}. Figure~\ref{2d2} embodies the collapse of MSD$_M(t)$ curves for the wide range of system sizes studied around the first transition point, obtained for $z_1 = 0.45 \pm 0.05$. At the intermediate regime of this plot, the curve is expected to decay as $\sim t^{\alpha-1}$. Numerically, we estimate the anomalous exponent to be $\alpha = 0.752 \pm 0.005$. 
Figure~\ref{2d3} now illustrates an analogous collapse of the curves for around the second transition point. This is attained by plotting $-\ln{\left(C_M(t)/C_M(0)\right)}/(L^{-z_2}t)$ as a function of $t/L^{z_2}$, where $z_{2} = 2.1665$ is set equal to the value for $z$ as reported by Nightingale and Bl\"ote~\cite{nightingale96}.

The intermediate regime for MSD$_M$ starts at time $\tau_1 \sim L^{z_1}$ at 
a value of $\langle (\Delta M)^2 \rangle \sim L^{2+z_1}$, then increases following a 
power-law mode with an exponent $\alpha$, until it reaches its saturation 
value $\sim L^{2+\gamma/\nu}$ at time $\tau_2 \sim L^{z_2}$. Assuming a single 
power-law function in the intermediate regime, the anomalous exponent is expected to be 
\begin{equation}
\label{eq:exporel}
\alpha = (\gamma/\nu-z_{1})/(z_{2} - z_{1}).
\end{equation}
Purely based on numerical findings, we speculate that $z_1 = 1/2$ and $\alpha = 3/4$; in that case, we obtain from Eq.~(\ref{eq:exporel}) that $z_2 = 13/6 = 2.1667$ in excellent agreement with the most accurate numerical estimates~\cite{nightingale96}.

To further corroborate on the main aftermath of our work, we undertook a parallel examination of the three-dimensional Ising ferromagnet. Analogously to the analysis sketched above for the two-dimensional Ising model, we obtained data collapses around the first and second crossover times. Figures~\ref{3d1} - \ref{3d3} below summarize our main findings: Fig.~\ref{3d1} exhibits the raw data, Fig.~\ref{3d2} suggests that $\text{MSD}_M(t)/(L^3 t)$ is a function of $t/L^{z_1}$ with $z_1 = 1.35\pm0.02$, and
Fig.\ref{3d3} that $-\ln{(\hat{C}_M(t))}/(L^{-z_{2}}t)$ is a function of $t/L^{z_2}$ with 
$z_{2} = 2.032 \pm 0.003$. Thus, as in two dimensions, the dynamical critical behavior features two crossover times characterized by two dynamic critical exponents. Additionally, the exponent of the intermediate anomalous diffusion $\alpha$ for the three-dimensional Ising ferromagnet is numerically found to be $0.90 \pm 0.02$. An overview of critical exponents reported in this manuscript is given in Tab.~\ref{p_wt}.

\begin{figure}[H]
	\includegraphics[width=8.5cm]{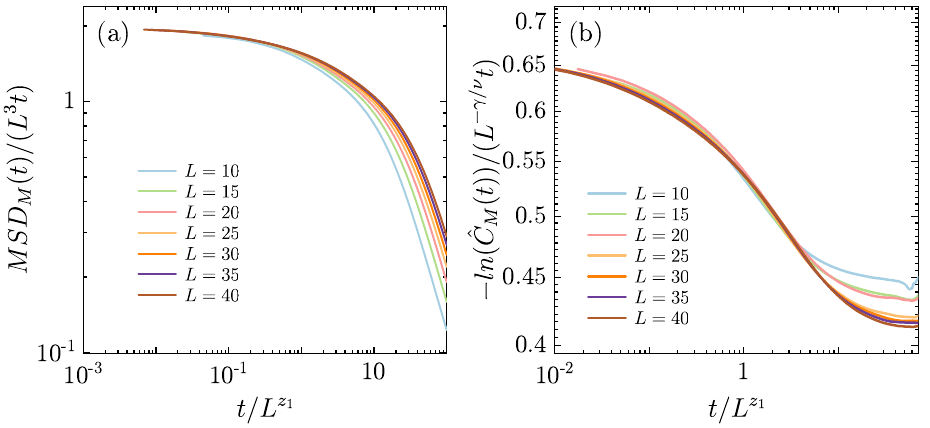}
	\caption{\label{3d2}
		Data collapse around the first crossover for the three-dimensional Ising model. (a) $\text{MSD}_M(t)$ collapse over various $L$, with a scaling form of $\text{MSD}_M(t)/(L^3 t) \sim t/L^{z_1}$, where the numerically found estimate for $z_1$ is $1.35 \pm 0.02$. $\text{MSD}_M(t)$ turns over from normal diffusion ($\sim L^3t$) to anomalous diffusion ($\sim L^{3+z_1-\alpha z_1}t^\alpha$) at $t=L^{z_1}$. (b) $-\ln{(\hat{C}_M(t))}$ collapse around $t = L^{z_{1}}$ with a scaling factor $L^{-\gamma/\nu}$ (note that $\gamma/\nu=1.9637$ in the three-dimensional Ising universality class~\cite{amit_book}). $\hat{C}_M(t)$ shifts from exponential to stretched exponential around $t = L^{z_{1}}$.}
\end{figure}

\begin{figure}[H]
	\includegraphics[width=8.5cm]{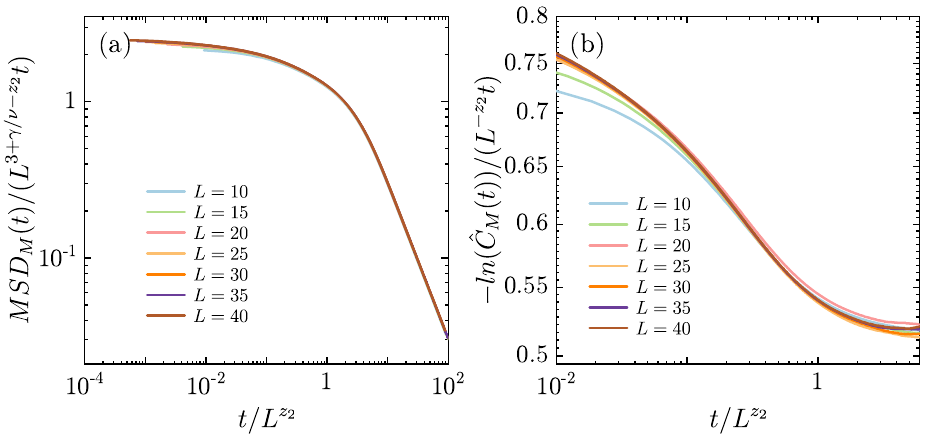}
	\caption{\label{3d3}
		Data collapse around the second crossover for the three-dimensional Ising model. (a) $\text{MSD}_M(t)$ collapse at $t\approx L^{z_{2}}$, with a scaling form of $MSD_M(t)/(L^\lambda t) \sim t/L^{z_{2}}$, the numerically found $\lambda$ and $z_2$ are $3+\gamma/\nu-z_{2}$ and $2.032 \pm 0.003$, respectively. $\text{MSD}_M(t)$ gradually transforms to saturation ($\sim L^{3+\gamma/\nu}$) from the anomalous diffusion ($\sim L^{3+z_1-\alpha z_1}t^\alpha$). (b) Data collapse for $-\ln{(\hat{C}_M(t))}$ around $t=L^{z_2}$, where the scaling factor $L^{-z_{2}}$ leads to excellent collapse. $\hat{C}_M(t)$ is expected to turn over from stretched exponential to exponential around $t = L^{z_{2}}$.}
\end{figure}

\begin{table}[H]
	\caption{\label{p_wt} A summary of critical exponents as reported in this manuscript for the two- (2D) and three-dimensional (3D) Ising ferromagnet. The last two columns refer to exact~\cite{landau_book} or high-precision~\cite{kos16} estimates of the critical exponents $\gamma$ and $\nu$ that have been used in the data collapse.}
	\begin{ruledtabular}
		\begin{tabular}{cccccc}  
			&$z_{1}$&$z_{2}$& $\alpha$&$\gamma$&$\nu$ \\ 
			\colrule 
			2D &$0.45(5)$&$2.1665(12)$& $0.752(5)$&$7/4$&$1$ \\ 
			3D &$1.35(2)$&$2.032(3)$& $0.90(2$)&$1.237075(10)$&$0.629971(4)$ \\ 
		\end{tabular}
	\end{ruledtabular}
\end{table}

\section{Summary and outlook}
\label{sec:summary}

We analyzed the results of extensive simulations of the two- and three-dimensional Ising model with Glauber dynamics. In particular, we scrutinized the mean-squared deviation and autocorrelation function of the magnetization, showcasing the existence of three dynamical regimes, separated by two crossover times at $\tau_1 \sim L^{z_1}$ and $\tau_2 \sim L^{z_2}$. In the short-time regime, the mean-squared deviation of the magnetization shows ordinary diffusive behavior and the autocorrelation function exponential decay. In the second intermediate regime the mean-squared deviation is characterized by anomalous diffusive behavior and the autocorrelation function decays as a stretched-exponential way. Finally, in the third late-time regime the mean-squared deviation saturates at a constant value while the autocorrelation function again decays exponentially. 

The second crossover to the exponential decay of the autocorrelation function has been extensively studied in the literature. Nightingale and Bl\"ote reported that this exponential decay sets in at a time determined by the dynamic critical exponent $z = 2.1665(12)$~\cite{nightingale96}; this is in agreement with our estimate $z_{2}$ at the second crossover. To the best of our knowledge, the first crossover has not yet been reported or was assumed to occur at some fixed time (i.e., $z_{1} = 0$) without substantiation. The simulations and analysis captured here clearly manifest the existence of this first crossover at a time governed by a new dynamic critical exponent $z_{1}$. We should stress here that earlier work on non-equilibrium dynamics has also suggested the presence of a new exponent $\theta$~\cite{janssen89} akin to the newly introduced exponent $z_1$ of the present work. The authors of Ref.~\cite{janssen89} considered a quench from a high temperature configuration with an initial magnetization $M(0)$ to the critical temperature $T_{\rm c}$; the exponent $\theta$ was introduced to describe the behavior
in the critical initial slip.

We also postulated a speculative argument about the crossover times at two dimensions. Purely on numerical grounds, we suspect the first crossover to correspond to a dynamic exponent $z_{1} = 1/2$, and the exponent of the anomalous diffusion to be $\alpha = 3/4$. In this case, we showed that the second crossover is governed by the exponent $z_{2} = 13/6$, in full agreement with the numerical result $z = 2.1665(12)$. At this stage, the development of a solid theoretical argument supporting the presence of the numerically observed first crossover and the relevant dynamic and anomalous diffusion exponent $z_1$ and $\alpha$ respectively is called for. 

To sum up, we hope that the relevance of our work will be twofold: (i) On the practical side, for obtaining statistically uncorrelated samples the proper sampling frequency should be set by the newly reported exponent $z_{1}$: the correlation between consecutive samples which are separated by (multiples of) $\tau_1 \sim L^{z_{1}}$ has decayed in a stretched-exponential way to a value which is as small as one would want. Hence, for obtaining statistically uncorrelated samples it is not necessary to sample with an interval scaling as $\tau_2$. (ii) On the theoretical side, the critical dynamical behavior of the Ising model with Glauber dynamics is much richer than reported till date featuring two distinct crossovers. Thus, if dynamic universality exists, it must also be much more substantial and needs further investigation. 

Closing, we would like to raise some motivational comments for future work. In a recent paper~\cite{zhong18} it was shown that the $\phi^4$ model with local dynamics appears to belong to the same dynamic universality class as the Ising model; this was done by probing numerically the dynamic critical exponent which was found to be $z = 2.17(3)$. If indeed this is the case, then also the exponent $z_{1}$ should apply to the $\phi^4$ model; see also Refs.~\cite{bloete95,bloete98,hasenbusch99} for extensive aspects on the dynamic Ising universality.
Furthermore, in Ref.~\cite{zhong19} the Ising model with Kawasaki dynamics was studied and the authors reported that the Fourier modes of the magnetization are in very close agreement with the dynamical eigenmodes, suggesting that $z = 4 - \eta = 15/4$. Investigating this aspect under the prism of the newly introduced exponent $z_{1}$ might be another intriguing continuation of our work~\cite{comment}. We plan to pursue these and other relevant open questions in the near future.

\begin{acknowledgments}
We would like to thank Peter Grassberger and Martin Hasenbusch for fruitful correspondence. We acknowledge the provision of computing time on the parallel computer cluster {\it Zeus} of Coventry University and T\"{U}B\.{I}TAK ULAKB\.{I}M (Turkish agency), High Performance and Grid Computing Center (TRUBA Resources). 
\end{acknowledgments}


\begin{thebibliography}{199}
\bibitem{comment} Of course the total magnetization (zero mode) is not a good observable in this case, as it is strictly conserved; so one has to study, for instance, the first Fourier mode of the magnetization.
\end{thebibliography}
\end{document}